\documentclass[a4paper]{article}

\usepackage{INTERSPEECH2019}
\usepackage{multirow}
\usepackage{float}
\usepackage{tabularx}

\makeindex

\title{FINE-GRAINED ROBUST PROSODY TRANSFER FOR SINGLE-SPEAKER NEURAL TEXT-TO-SPEECH}
\name{Viacheslav Klimkov, Srikanth Ronanki, Jonas Rohnke, Thomas Drugman}
\address{
  Amazon Research, Cambridge}
\email{vklimkov,ronanks,rohnj,drugman@amazon.com}

\begin{document}

\maketitle
\begin{abstract}

We present a neural text-to-speech system for fine-grained prosody transfer from one speaker to another. Conventional approaches for end-to-end prosody transfer typically use either fixed-dimensional or variable-length prosody embedding via a secondary attention to encode the reference signal. However, when trained on a single-speaker dataset, the conventional prosody transfer systems are not robust enough to speaker variability, especially in the case of a reference signal coming from an unseen speaker. Therefore, we propose decoupling of the reference signal alignment from the overall system. For this purpose, we pre-compute phoneme-level time stamps and use them to aggregate prosodic features per phoneme, injecting them into a sequence-to-sequence text-to-speech system. We incorporate a variational auto-encoder to further enhance the latent representation of prosody embeddings. We show that our proposed approach is significantly more stable and achieves reliable prosody transplantation from an unseen speaker. We also propose a solution to the use case in which the transcription of the reference signal is absent. We evaluate all our proposed methods using both objective and subjective listening tests.

\end{abstract}
\noindent\textbf{Index Terms}: Neural text-to-speech, sequence-to-sequence, prosody transfer.

\section{Introduction}

Neural text-to-speech (NTTS) \let\thefootnote\relax\footnotetext{Paper accepted for Interspeech 2019} methods significantly boosted the overall naturalness of synthetic speech \cite{sotelo2017char2wav, shen2018natural, li2018close} while allowing to build much more flexible synthesis systems \cite{prateek2019style, skerry2018towards, gibiansky2017deep}. As `neural text-to-speech', we here refer to a sequence-to-sequence (seq2seq) model predicting mel-spectrograms, followed by a neural vocoder as proposed in Tacotron2 \cite{shen2018natural}. There have been several applications of NTTS for style \cite{wang2018style, prateek2019style}, prosody \cite{skerry2018towards} and speaker \cite{gibiansky2017deep} control. Fine-grained Prosody Transfer (PT) is another possible extension of NTTS, allowing precise control over temporally dependent prosodic structures, such as phrase breaks, emphasis, prominence, etc.
A particular shortcoming of fine-grained PT approaches is that they are quite limited in their ability to generalize well for long utterances and hence prevents the use of synthesized speech in otherwise appealing applications such as voice imitation for audiobooks. In this paper, we focus on seq2seq models for achieving robust PT from one speaker to another under various training conditions.

\subsection{Relation to prior work}

End-to-end PT has recently gained interest in a number of studies \cite{skerry2018towards, wang2018style, zhang2018learning}. The vast majority of them focus on sentence-level prosody embeddings. These work well for transplanting general voice quality, overall duration and pitch of the sentence but the fine-grained prosody control is rather limited. 

In \cite{skerry2018towards}, an extension to the Tacotron architecture \cite{wang2017tacotron} is proposed, which compresses the prosody of a whole utterance into a fixed-dimension embedding, losing temporal information. For example, phrasing structure, emphasis and accents are not transplanted properly. Global Style Tokens (GSTs) are introduced in \cite{wang2018style} to encode different speaking styles, when jointly trained within Tacotron. Similar to \cite{skerry2018towards}, this approach uses a reference encoder to guide GST weights in order to synthesize a phrase in the prosodic style of a reference speech. In \cite{zhang2018learning}, a Variational Auto Encoder (VAE) is proposed in an end-to-end speech synthesis model for style control and transfer. All these approaches use a reference encoder and a fixed-size embedding for transferring style at utterance level, but do not achieve fine-grained control or transfer of prosody. 

Another challenge for end-to-end PT is speaker perturbation with the ultimate goal of transplanting prosody from arbitrary speakers. This will allow to synthesize speech with natural prosody from a given speaker, using a reference signal from any other speaker. In a recent study \cite{lee2018robust}, a secondary attention module within the seq2seq framework is proposed  to overcome some of these limitations. They also propose to use normalized embeddings to address speaker perturbations, and variable-length prosody embeddings to enable fine-grained prosody control. However, the secondary attention used in \cite{lee2018robust} to map the reference signal and text, does not generalize well when the model is trained on a single speaker and tested with reference signal from unseen speaker.

In general, we identify two potential shortcomings of the aforementioned approaches. First, fine-grained PT is still not robust enough to use it in audiobook applications. Secondly, the transfer of prosody for NTTS in the absence of input text has never been attempted, to the best of our knowledge.

\subsection{Novelty of this work}

We propose a novel robust approach for fine-grained PT and evaluate it for a single-speaker scenario which is a typical case for TTS datasets. We study if it helps to decouple the acoustic feature generation from the task of finding the correspondence of text and reference signal. We achieve this by performing forced alignment \cite{ljolje1991automatic} (a.k.a. phoneme segmentation) of the reference signal with a generic acoustic model beforehand, aggregating the prosodic features on a phoneme-basis and injecting them into the seq2seq system. In addition, we explore a variational auto-encoder approach as proposed in \cite{zhang2018learning} to enhance the interpolation ability of the prosody transplantation framework. 

A vast majority of TTS datasets are created for unit-selection technology and are not expressive (i.e. they have a flat prosody). Therefore, utilizing non-expressive data is a valuable practical research direction. We study the effect of dataset expressivity and the reduction of PT capabilities of a model built with such a dataset. Another practical limitation of PT is also explored, where reliable transcript for the speech to be resynthesized is not available. We perform prosody transfer in the absence of input text, where the output of an Automatic Speech Recognition (ASR) model is directly fed to the speech synthesis module together with reference audio.

The paper is organized as follows: Section 2 describes the conventional model for NTTS and a secondary attention module for end-to-end prosody transplantation framework; Section 3 introduces the proposed PT using an aggregated reference signal and a variational autoencoder. We evaluate our proposed approach for PT between two speakers under various training conditions in Section 4. Finally, in Section 5, we summarize our conclusions.

\section{Baseline model}
\label{sec:baseline}

\begin{figure}[t]                                                            
\hspace*{-5mm}                                     
\centering                                                             
\includegraphics[width=90mm]{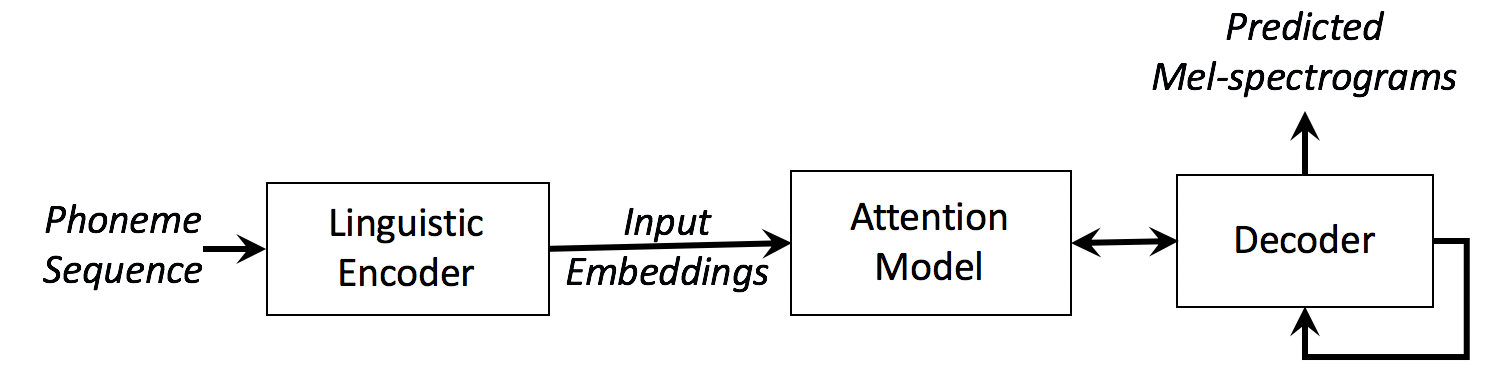}
\vspace{-4mm}
\caption {Schematic diagram of a seq2seq Neural TTS} 
\label{fig:baseline}
\end{figure}

The system architecture for our baseline NTTS model follows that of Tacotron2 \cite{shen2018natural}, with minor changes.
First, a seq2seq acoustic model predicts mel-spectrograms from a sequence of phoneme-level linguistic inputs. Then a speaker-independent neural vocoder converts the mel-spectrograms into a high-fidelity audio waveform \cite{lorenzo2018robust}. 

The schematic diagram of a general seq2seq Neural TTS is presented in Figure \ref{fig:baseline}. The seq2seq model consists of an encoder, auto-regressive decoder and an attention module which finds the correspondence between textual and acoustic representation of speech. Similar to Tacotron2, our encoder consists of a convolution stack and a bi-LSTM layer. As an input to the encoder, we use phoneme identities to speed up the training. The decoder is a stack of LSTM layers and a locally-sensitive version of the attention mechanism is utilized for training the network. The fine details of our input phoneme representation and the hyper-parameters used for training a seq2seq model are explained in detail in our previous work \cite{latorre2018effect}. 

\subsection{Secondary attention for PT}
\label{ssec:sec_att}

\begin{figure}[b]                                                            
\hspace*{-5mm}                                     
\centering                                                             
\includegraphics[width=90mm]{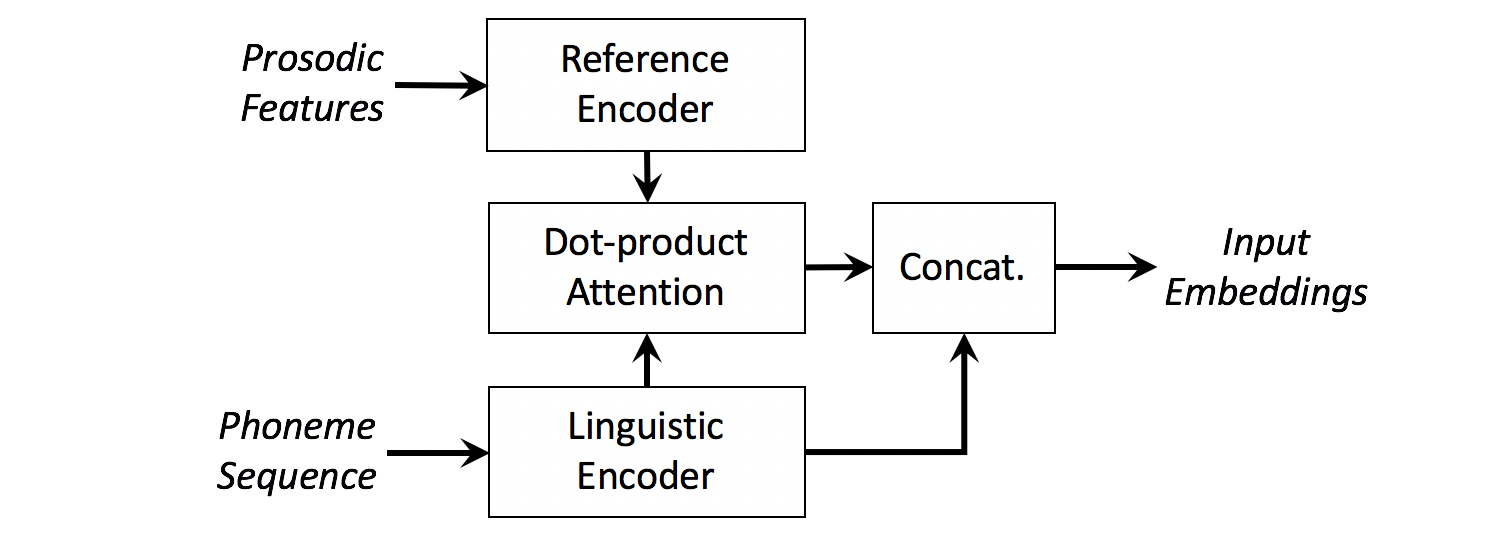}
\vspace{-2mm}
\caption {Secondary attention for PT} 
\label{fig:sec_att}
\vspace{-4mm}
\end{figure}

The use of prosody embedding  for NTTS was first introduced in \cite{skerry2018towards} by encoding a reference speech of an arbitrary speaker into a fixed-length learned representation using a reference encoder. The prosody embedding from the reference encoder is then broadcast-concatenated with the linguistic encoder representation to form a sequence of encoder embeddings. The use of fixed-length prosody embedding does not allow fine-grained prosody transplantation. To overcome this, variable-length prosody embedding via a secondary attention \cite{lee2018robust} is used as a conditional input along with embeddings from linguistic encoder as shown in Figure \ref{fig:sec_att}. We consider this approach as a baseline for our PT experiments in Section 4. Mel-spectrograms were used as an input to the reference encoder in the original study. That required per-speaker normalization of prosody embeddings within the network. In contrast, we used prosodic features that are easy to normalize beforehand: $pitch$ and $power$.

\section{Proposed approach for PT}

In this section, we first propose the use of aggregated reference for PT. Then we show the application of VAE for better generalization towards unseen speakers.

\subsection{Aggregated reference for PT} 
\label{ssec:Agg}

In case of single-speaker training dataset, the approach from Section \ref{ssec:sec_att} suffers from instabilities of the secondary attention.
For long utterances, the secondary attention is quite unstable and the PT does not occur.
Furthermore, prosody-only features do not contain sufficient information to reliably perform the alignment of reference signal.

\begin{figure}[t]                                                            
\vspace{-3mm}      
\hspace*{-3mm}                        
\centering                                                             
\includegraphics[width=90mm]{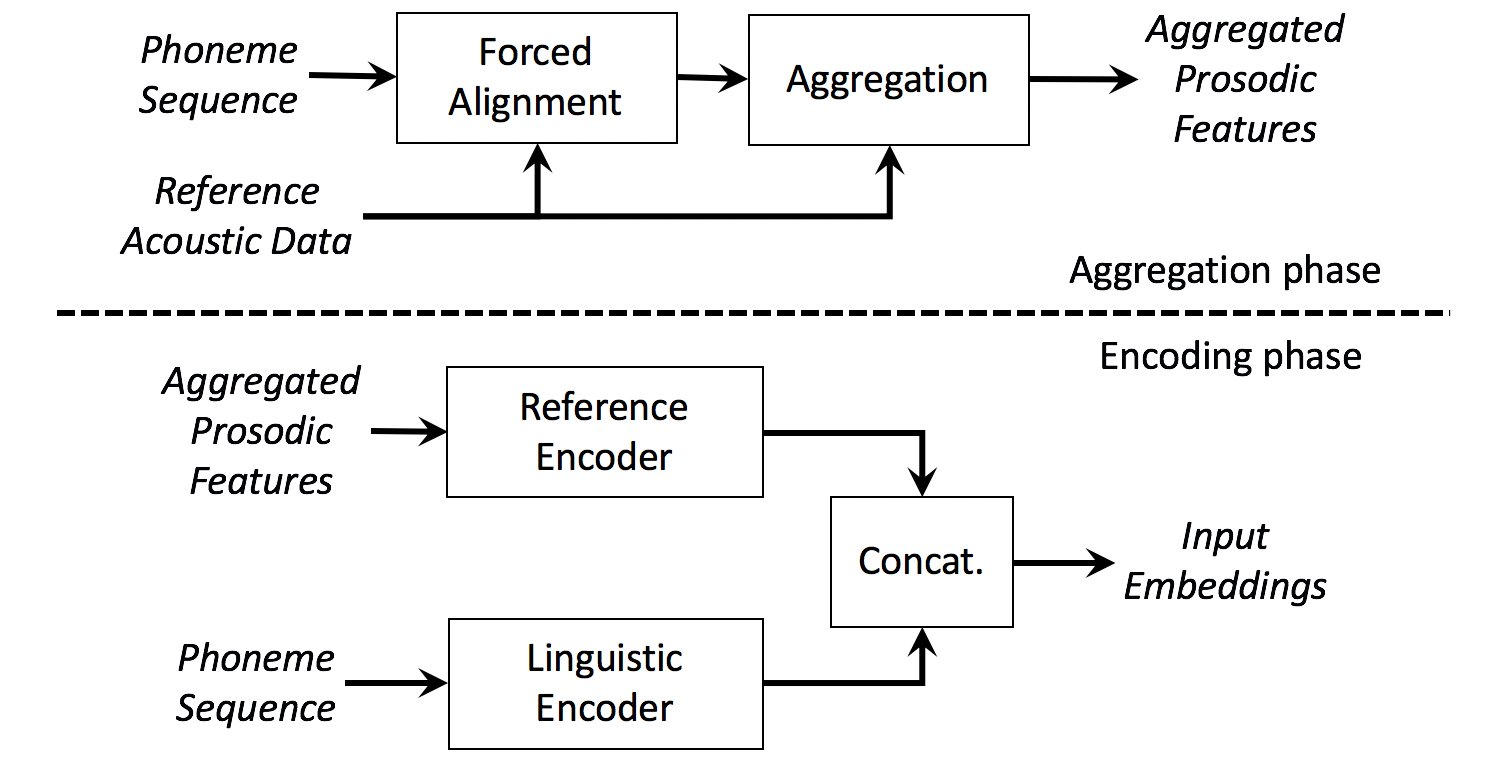}
\vspace{-5mm}
\caption {PT using aggregated reference signal} 
\label{fig:agg}
\vspace{-6mm}
\end{figure}

Since TTS models are usually trained on a single-speaker dataset, it is worth separating the alignment of the reference signal from the actual speech synthesis. 
The proposed approach is shown in Figure \ref{fig:agg}. First, the \textit{Aggregation phase} is performed. It includes forced alignment of the reference signal and uses the resulting phoneme boundary information to aggregate prosodic features per phoneme by performing averaging of frame-level features.
During the \textit{Encoding phase}, aggregated prosodic features are passed through the reference encoder of the same architecture as in Section 2, and are concatenated with linguistic encoder outputs.

A 7-dimensional prosodic representation was used for each phoneme, which is composed of mean $F0$ and $mgc0$ for each of 3 phoneme states, and phoneme duration. All features are normalized by mean and variance per speaker, duration is additionally normalized per phoneme identity.

\subsection{Aggregated variational reference for PT} \label{ssec:AggVae}

Variational Auto Encoders (VAE) were first introduced in \cite{kingma2013auto} and have been used in the context of end-to-end speech synthesis (for e.g., \cite{zhang2018learning,akuzawa2018expressive}).
Using VAE in the context of PT forces the latent space of prosody embeddings to be uniform and continuous, which further improves the stability of transfer from unseen speakers.

\begin{figure}[H] 
\vspace{-3mm}                                                             
\hspace*{-5mm}                                     
\centering                                                             
\includegraphics[width=90mm]{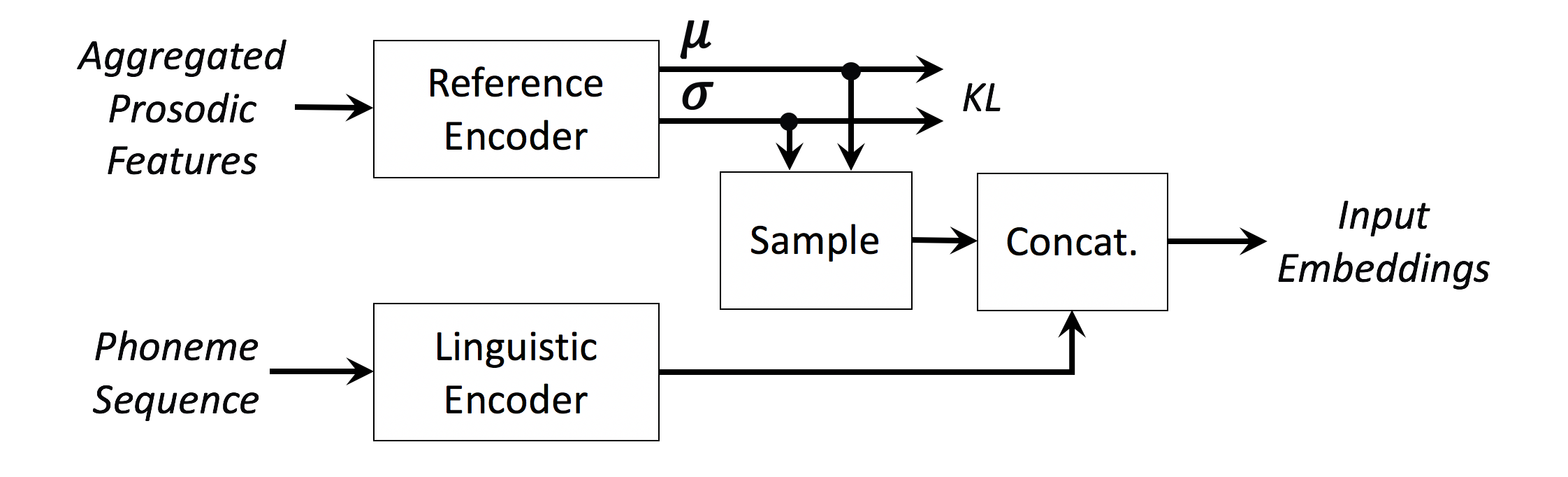}
\caption {PT using aggregated reference and VAE} 
\vspace{-1mm}
\label{fig:agg_vae}
\end{figure}

Figure \ref{fig:agg_vae} shows the modifications to the \textit{Encoding phase} - the aggregation phase remains unchanged. The reference encoder predicts distributions (mean $\mu$ and standard deviation $\sigma$) of prosody embedding for each phoneme, from which we sample prosody embeddings. Predicted distributions are used in a Kullback-Leibler (KL) divergence loss component, that constrains the latent space to be uniform and continuous.

During training, the convergence speed of the KL loss usually surpasses that of the reconstruction loss. 
This leads to the well-documented KL loss collapse. Similar to \cite{zhang2018learning}, we found that introducing KL annealing and scheduling helps to avoid this problem. In our experiments we linearly increase a scaling factor for the KL loss from 0 to 1 between iterations 25k and 150k. In addition, we only take the KL loss into account every 200 training steps.

\section{Experiments and results}

\subsection{Data} \label{ssec:Data}
We conducted experiments on an internal US English dataset of audiobook recordings. The training dataset consists of 20 hours of recordings from 4 non-fiction audiobooks, read in an expressive style by a female speaker. For the results presented in section \ref{ssec:EvalRes}, two sets of 50 utterances were used. The first one comes from held-out books for the same speaker (SS) who narrated the training data; the second one is from an unseen speaker (US) narrating unseen books in a similar style. The unseen speaker is also a US English female voice. Speakers were selected arbitrarily without taking voice similarity into account.

\subsection{Evaluation protocol} 
\label{ssec:Eval}

Four systems were evaluated: 1) \textit{Base} - baseline NTTS without PT; 2) \textit{Sec\_Att} - PT using secondary attention (section \ref{ssec:sec_att}); 3) \textit{Agg} - PT using aggregated reference (Section \ref{ssec:Agg}); 4) \textit{Agg\_VAE} - PT using aggregated reference with VAE (Section \ref{ssec:AggVae}).

\vspace{2mm}
\noindent \textbf{Objective metrics:} Objective metrics are used to compare acoustic parameters extracted from synthesized audio and natural recordings. We only focus on comparing $F0$ values. For all comparisons, we use Dynamic Time Warping (DTW) \cite{bellman1959adaptive} to match the predicted to the reference sequence length. Since mel-cepstral features are well-suited for DTW alignment, the path obtained from mel-cepstral features is used to align $F0$ as well.
We report $F0$ root mean square error (RMSE), correlation and F0 frame error (FFE) \cite{chu2009reducing, drugman2011joint}. 

\vspace{2mm}
\noindent \textbf{Subjective tests:} The systems were evaluated subjectively using a MUSHRA test \cite{itu20031534}. 24 native English speakers were presented with reference audio and samples from the systems in random order side-by-side, and were asked to \textit{``Rate the systems in terms of their naturalness and prosody transfer. 100 means a very natural speech, prosody matches exactly with reference audio and 0 means a very unnatural speech and doesn't match at all''}. A total of 96 utterances were used for testing and the test is balanced in such a way that each listener scored 32 out of 96 (16 from same speaker and another 16 from unseen speaker). Actual recordings are used as an upper anchor. The significance of the MUSHRA results was analyzed with a Wilcoxon signed-rank test and a standard t-test, both with Bonferroni-Holm correction applied \cite{clark2007statistical}.

\subsection{Evaluation results} 
\label{ssec:EvalRes}

Objective metrics are reported in Table \ref{table:obj}.
As expected, the baseline system achieves a low F0 correlation.
PT with secondary attention performs quite good, but introduces a significant degradation (all three metrics) if an unseen speaker is used as prosody reference. The standard deviation of F0 correlation in the $Sec\_Att$ approach illustrates the effect of its instability for unseen speakers. Our proposed solution is also affected but to a lesser extent.  Introducing VAE to the aggregated approach slightly improves all the objective metrics.

\begin{table}[t]  
\caption{Objective metrics for different PT approaches. Labels for reference: Same speaker (SS), Unseen speaker (US)}
\vspace{-2mm}
\begin{tabular}{|l|c|c|c|c|}
\hline
\textbf{Model}            & \textbf{Ref.} & \textbf{RMSE (Hz)} & \textbf{CORR} & \textbf{FFE (\%)} \\ \hline
\multirow{2}{*}{Base}  & SS  & 42.7 $\pm$ 8.1 & 0.45 $\pm$ 0.16  & 29.72            \\ \cline{2-5} 
                          & US & 36.3 $\pm$ 8.7 & 0.57 $\pm$ 0.14 & 31.05 \\ \hline
\multirow{2}{*}{Sec\_Att} & SS & 18.9 $\pm$ 7.4 & 0.86 $\pm$ 0.09 & 9.79             \\ \cline{2-5} 
                          & US & 24.1 $\pm$ 13.1 & 0.78 $\pm$ 0.16 & 20.18 \\ \hline
\multirow{2}{*}{Agg} & SS & 16.4 $\pm$ 7.4 & 0.89 $\pm$ 0.08 & 9.09 \\ \cline{2-5} 
                          & US & 20.5 $\pm$ 7.8 & 0.84 $\pm$ 0.09 & 15.05            \\ \hline
\multirow{2}{*}{Agg\_VAE} & SS & 16.4 $\pm$ 6.8 & 0.89 $\pm$ 0.08 & 8.93 \\ \cline{2-5} 
                          & US & 20.1 $\pm$ 7.2 & 0.85 $\pm$ 0.09 & 14.98            \\ \hline
\end{tabular}
\label{table:obj}
\vspace{-5mm}
\end{table}

\begin{table}[H]  
\caption{MUSHRA medians for different PT approaches. Labels for reference: Same speaker (SS), Unseen speaker (US)}
\vspace{-2mm}
\centering
\begin{tabular}{|c|c|c|c|c|}
\hline
\textbf{Ref.} & \textbf{Base} & \textbf{Sec\_Att} & \textbf{Agg} & \textbf{Agg\_VAE}   \\ \hline
SS         & 30.0       & 72.5            & 77.0         & 78.0     \\ \hline
US         & 30.0       & 59.0            & 66.0         & 70.0     \\ \hline
\end{tabular}
\label{table:mushra_median}
\vspace{-3mm}
\end{table}

\begin{figure}[H]                                                         
\hspace*{-5mm}                                     
\centering                                                             
\includegraphics[width=70mm]{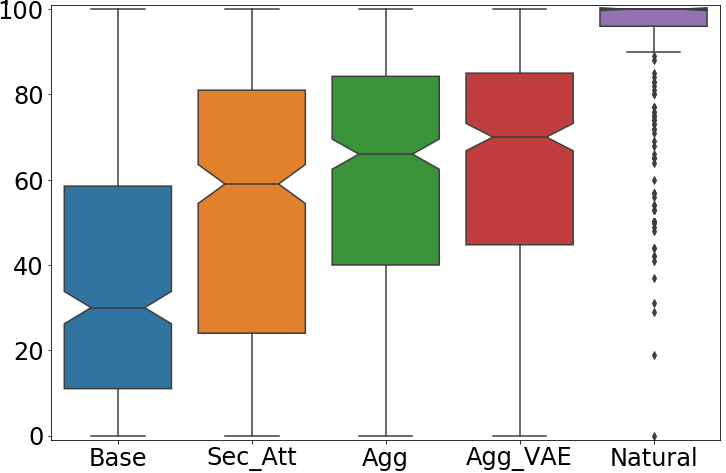}
\vspace{-1mm}
\caption {Subjective listeners ratings from a MUSHRA test.} \label{fig:k1}
\end{figure}

The MUSHRA results for subjective evaluation of different PT approaches are summarized in Table \ref{table:mushra_median}. Significant improvement over the secondary attention method is observed by using the proposed aggregated technique both in cases of PT from seen and unseen speaker ($p$-$value<0.01$ in pair-wise two-sided t-test and $p$-$value<0.05$ in Wilcoxon signed-rank test). VAE introduces additional improvement but not significant in case of same speaker. We present listener ratings with boxplots for the more challenging scenario of PT from an unseen speaker in Figure \ref{fig:k1}. All comparisons from this figure are statistically significant ($p$-$value<0.01$ in t-test).

\vspace{-3mm}

Per-case analysis of listeners judgements revealed that when the secondary attention was operating properly, there is only a small difference between $Agg$ and $Sec\_Att$. The differences are more pronounced when the secondary attention fails to assign correspondence between the reference signal and linguistic embeddings. In such cases, the synthesis had poor segmental quality and the PT did not happen and this is reflected in the boxplot in form of a stretched section below the median for $Sec\_Att$.

\subsection{PT model built with neutral dataset} 
\label{ssec:NeutralDataset}

Another important factor of prosody controllability is the data used for building the speech synthesis system. The majority of datasets for speech synthesis systems are recorded in a neutral speaking style. We trained the proposed system (Agg\_VAE) from Section \ref{ssec:AggVae} on an internal US English female speaker neutral dataset (ND) comparable in size to the expressive dataset used in previous experiments. 

\begin{table}[t]
\caption{Objective metrics for PT with (WT) and without (WOT) text transcription, trained on neutral dataset (ND)}
\vspace{-2mm}
\begin{tabular}{|c|c|c|c|c|c|}
\hline
\textbf{Data} & \textbf{Text} & \textbf{Ref.} & \textbf{RMSE (Hz)} & \textbf{CORR} & \textbf{FFE (\%)} \\ \hline
ND & WT & US                 & 41.9           & 0.79            & 52.7            \\ \hline
ND & WOT & US      & 45.5           & 0.75            & 55.7            \\ \hline
\end{tabular}
\label{table:ctc}
\vspace{-5mm}
\end{table}

The results of the evaluation in terms of PT from an unseen speaker are presented in the first row of Table \ref{table:ctc}.
There is a substantial drop in PT capabilities compared to the results from Table \ref{table:obj}. 
We speculate that the reason for this is that the same value of the normalized reference signal corresponds to very different levels of expressivity for the speaker observed during training and the unseen speaker during inference. In addition, the rapid dynamics of expressive prosody are mapped to much smaller changes observed in the training data.
Despite a lower expressivity, the PT is still effective, as reflected by the $F0$ correlation.
The PT with aggregated prosody is robust enough to consistently transplant prosody without instabilities even   when there is a significant discrepancy between the reference signal observed during the training and inference.

\subsection{PT in absence of text transcription}

The need for accurate text transcription of the reference signal is a drawback of NTTS with PT capabilities. To alleviate this, we propose text-less prosody transfer using just audio.

\begin{figure}[b]      
\hspace*{-5mm}                                                                                        
\centering                                                             
\includegraphics[width=90mm]{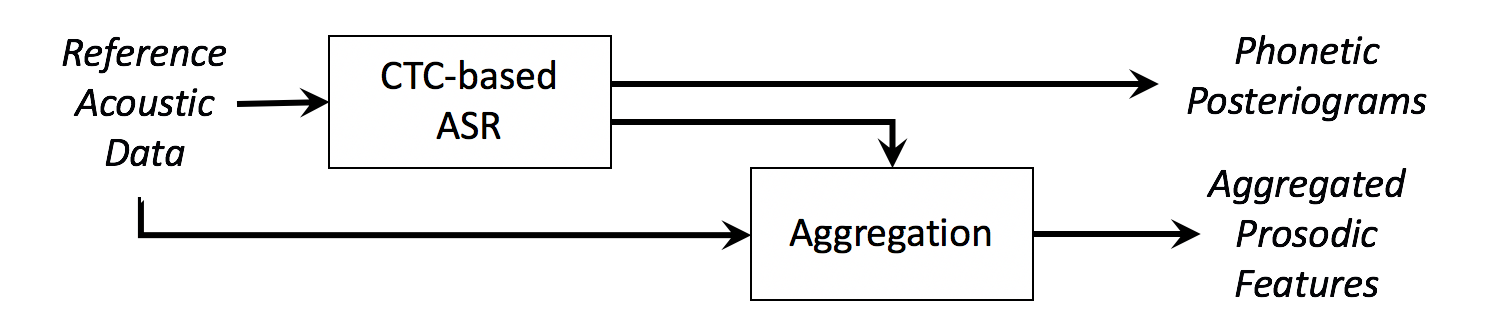}
\caption {Aggregation phase in absence of transcripts} 
\label{fig:ctc}
\vspace{-3mm}
\end{figure}

Studies of voice-conversion on non-parallel data \cite{sun2016phonetic,liu2018voice} use a speech recognition module to generate phonetic posteriorgrams which are used as an input to speech synthesis module. We expand this approach by using posteriorgrams together with prosodic features to achieve PT. The work flow of the aggregation phase is depicted in Figure \ref{fig:ctc}. During the \textit{Encoding phase}, we use the \textit{Phonetic Posteriorgrams} instead of the (unavailable) \textit{Phoneme sequences}. They are informative about particular realizations of phonemes with respect to acoustics. They also help to recover from recognition errors that are inevitable for unconstrained phoneme recognition.

We use a Connectionist Temporal Classification (CTC) based end-to-end ASR system as in \cite{hannun2014deep} to predict phoneme identities for given audio. As training data, we use a combination of \cite{panayotov2015librispeech} and \cite{commonvoice2013}. 
For each audio file in the TTS training data, the sequence of non-blank phonemes is composed and used as input to the TTS model which is trained on audio-only dataset from Section \ref{ssec:NeutralDataset}. 
In case of a blank phoneme spanning a sufficient amount of time ($>$200 ms), we interpret it as a ``pau'', which was also included into the phoneme sequence. 
Assuming that non-blank phoneme labels are emitted in the stable part of the phone, the prosodic features from the corresponding audio regions are used as a reference signal to achieve PT.
The distance to previous and subsequent phonemes is also included into the reference signal to represent the duration of the given phoneme in the reference signal. The same normalization strategies as in Section \ref{ssec:Agg} are applied.

\begin{figure}[t]       
\hspace*{-6mm}                                                                                     
\includegraphics[width=90mm]{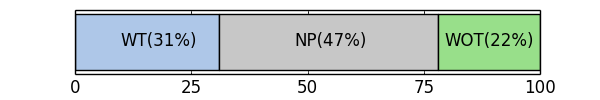}
\caption {Results of the subjective evaluation of text-less PT. WT denotes the system trained with text from Section \ref{ssec:NeutralDataset}, NP denotes no preference, and WOT denotes the system trained without text, using phonetic posteriograms.} 
\label{fig:subjctc}
\vspace{-3mm}
\end{figure}

The objective performance of PT from an unseen speaker is shown in the second row of Table \ref{table:ctc}. The results show a slight degradation compared to PT with text.
As a subjective evaluation, a preference test was carried out using Amazon Mechanical Turk platform. 10 native English speakers were asked to \textit{"Select which audio sounds more natural"} for 50 pairs of audio samples created with ($WT$) and without ($WOT$) the use of text. The results of the preference test are shown on Figure \ref{fig:subjctc}.

There is a statistically insignificant ($p$-$value>0.01$) preference towards the system that utilizes text transcription.
A manual inspection of cases when the $WT$ system was preferred revealed that occasionally, despite using posteriograms rather than phoneme identities, the errors in speech recognition lead to incorrect pronunciations which affect the listeners judgments.
PT was not affected per se and both approaches rendered very similar prosody patterns.
We show that PT without any transcript is achievable with almost the same level of quality.

\section{Conclusions}

In this work, we have introduced a neural text-to-speech approach for fine-grained prosody transfer. The proposed approach aligns a reference signal with a phoneme sequence for synthesis beforehand and is robust for prosody transfer from an unseen speaker when trained on a single-speaker dataset. We have also demonstrated that additional improvements can be achieved by incorporating a variational autoencoder. Comparing the proposed approaches to prosody transfer using a secondary attention shows consistent improvement of signal quality and prosody transfer capabilities as well as a significant improvement of stability for the edge cases in both objective and subjective evaluations.

We studied the effect of using a neutral dataset for building a TTS system with the possibility of prosody transfer and to measure the degradation of expressivity. Finally, we proposed and evaluated an approach for text-less prosody transfer which achieves a statistically insignificant degradation of quality in the absence of text transcription.

\bibliographystyle{IEEEtran}
\bibliography{references}

\begin{thebibliography}{10}
\providecommand{\url}[1]{#1}
\csname url@samestyle\endcsname
\providecommand{\newblock}{\relax}
\providecommand{\bibinfo}[2]{#2}
\providecommand{\BIBentrySTDinterwordspacing}{\spaceskip=0pt\relax}
\providecommand{\BIBentryALTinterwordstretchfactor}{4}
\providecommand{\BIBentryALTinterwordspacing}{\spaceskip=\fontdimen2\font plus
\BIBentryALTinterwordstretchfactor\fontdimen3\font minus
  \fontdimen4\font\relax}
\providecommand{\BIBforeignlanguage}[2]{{%
\expandafter\ifx\csname l@#1\endcsname\relax
\typeout{** WARNING: IEEEtran.bst: No hyphenation pattern has been}%
\typeout{** loaded for the language `#1'. Using the pattern for}%
\typeout{** the default language instead.}%
\else
\language=\csname l@#1\endcsname
\fi
#2}}
\providecommand{\BIBdecl}{\relax}
\BIBdecl

\bibitem{sotelo2017char2wav}
J.~Sotelo, S.~Mehri, K.~Kumar, J.~F. Santos, K.~Kastner, A.~Courville, and
  Y.~Bengio, ``Char2wav: End-to-end speech synthesis,'' in \emph{ICLR 2017
  workshop}, 2017.

\bibitem{shen2018natural}
J.~Shen, R.~Pang, R.~J. Weiss, M.~Schuster, N.~Jaitly, Z.~Yang, Z.~Chen,
  Y.~Zhang, Y.~Wang, R.~Skerrv-Ryan \emph{et~al.}, ``Natural tts synthesis by
  conditioning wavenet on mel spectrogram predictions,'' in \emph{Proc.
  {ICASSP}}, 2018, pp. 4779--4783.

\bibitem{li2018close}
N.~Li, S.~Liu, Y.~Liu, S.~Zhao, M.~Liu, and M.~Zhou, ``Neural speech synthesis
  with transformer network,'' \emph{arXiv preprint arXiv:1809.08895}, 2018.

\bibitem{prateek2019style}
N.~Prateek, M.~Lajszczak, R.~Barra-Chicote, T.~Drugman, J.~Lorenzo-Trueba,
  T.~Merritt, S.~Ronanki, and T.~Wood, ``In other news: A bi-style
  text-to-speech model for synthesizing newscaster voice with limited data,''
  \emph{Accepted for NAACL}, 2019.

\bibitem{skerry2018towards}
R.~Skerry-Ryan, E.~Battenberg, Y.~Xiao, Y.~Wang, D.~Stanton, J.~Shor, R.~Weiss,
  R.~Clark, and R.~A. Saurous, ``Towards end-to-end prosody transfer for
  expressive speech synthesis with tacotron,'' in \emph{Proc. {ICML}}, 2018,
  pp. 4700--4709.

\bibitem{gibiansky2017deep}
A.~Gibiansky, S.~Arik, G.~Diamos, J.~Miller, K.~Peng, W.~Ping, J.~Raiman, and
  Y.~Zhou, ``Deep voice 2: Multi-speaker neural text-to-speech,'' in
  \emph{Proc. {NIPS}}, 2017, pp. 2962--2970.

\bibitem{wang2018style}
Y.~Wang, D.~Stanton, Y.~Zhang, R.~Skerry-Ryan, E.~Battenberg, J.~Shor, Y.~Xiao,
  F.~Ren, Y.~Jia, and R.~A. Saurous, ``Style tokens: Unsupervised style
  modeling, control and transfer in end-to-end speech synthesis,'' \emph{arXiv
  preprint arXiv:1803.09017}, 2018.

\bibitem{zhang2018learning}
Y.-J. Zhang, S.~Pan, L.~He, and Z.-H. Ling, ``Learning latent representations
  for style control and transfer in end-to-end speech synthesis,'' \emph{arXiv
  preprint arXiv:1812.04342}, 2018.

\bibitem{wang2017tacotron}
Y.~Wang, R.~Skerry-Ryan, D.~Stanton, Y.~Wu, R.~J. Weiss, N.~Jaitly, Z.~Yang,
  Y.~Xiao, Z.~Chen, S.~Bengio \emph{et~al.}, ``Tacotron: Towards end-to-end
  speech synthesis,'' in \emph{Proc. Interspeech}, 2017, pp. 4006--4010.

\bibitem{lee2018robust}
Y.~Lee and T.~Kim, ``Robust and fine-grained prosody control of end-to-end
  speech synthesis,'' \emph{arXiv preprint arXiv:1811.02122}, 2018.

\bibitem{ljolje1991automatic}
A.~Ljolje and M.~Riley, ``Automatic segmentation and labeling of speech,'' in
  \emph{Proc. {ICASSP}}, 1991, pp. 473--476.

\bibitem{lorenzo2018robust}
J.~Lorenzo-Trueba, T.~Drugman, J.~Latorre, T.~Merritt, B.~Putrycz, and
  R.~Barra-Chicote, ``Robust universal neural vocoding,'' \emph{arXiv preprint
  arXiv:1811.06292}, 2018.

\bibitem{latorre2018effect}
J.~Latorre, J.~Lachowicz, J.~Lorenzo-Trueba, T.~Merritt, T.~Drugman,
  S.~Ronanki, and V.~Klimkov, ``Effect of data reduction on
  sequence-to-sequence neural tts,'' in \emph{Proc. {ICASSP}}, 2019.

\bibitem{kingma2013auto}
D.~P. Kingma and M.~Welling, ``Auto-encoding variational bayes,'' \emph{arXiv
  preprint arXiv:1312.6114}, 2013.

\bibitem{akuzawa2018expressive}
K.~Akuzawa, Y.~Iwasawa, and Y.~Matsuo, ``Expressive speech synthesis via
  modeling expressions with variational autoencoder,'' \emph{arXiv preprint
  arXiv:1804.02135}, 2018.

\bibitem{bellman1959adaptive}
R.~Bellman and R.~Kalaba, ``On adaptive control processes,'' \emph{IRE
  Transactions on Automatic Control}, vol.~4, no.~2, pp. 1--9, 1959.

\bibitem{chu2009reducing}
W.~Chu and A.~Alwan, ``Reducing f0 frame error of f0 tracking algorithms under
  noisy conditions with an unvoiced/voiced classification frontend,'' in
  \emph{Proc. {ICASSP}}, 2009, pp. 3969--3972.

\bibitem{drugman2011joint}
T.~Drugman and A.~Alwan, ``Joint robust voicing detection and pitch estimation
  based on residual harmonics,'' in \emph{Proc. Interspeech}, 2011, pp.
  1973--1976.

\bibitem{itu20031534}
R.~B. ITU-R, ``1534-1,“method for the subjective assessment of intermediate
  quality levels of coding systems (mushra)”,'' \emph{International
  Telecommunication Union}, 2003.

\bibitem{clark2007statistical}
R.~A. Clark, M.~Podsiadlo, M.~Fraser, C.~Mayo, and S.~King, ``Statistical
  analysis of the blizzard challenge 2007 listening test results,'' in
  \emph{Proc. Blizzard Challenge Workshop}, vol. 2007, 2007.

\bibitem{sun2016phonetic}
L.~Sun, K.~Li, H.~Wang, S.~Kang, and H.~Meng, ``Phonetic posteriorgrams for
  many-to-one voice conversion without parallel data training,'' in
  \emph{International Conference on Multimedia and Expo (ICME)}, 2016, pp.
  1--6.

\bibitem{liu2018voice}
S.~Liu, J.~Zhong, L.~Sun, X.~Wu, X.~Liu, and H.~Meng, ``Voice conversion across
  arbitrary speakers based on a single target-speaker utterance,'' in
  \emph{Proc. Interspeech}, 2018, pp. 496--500.

\bibitem{hannun2014deep}
A.~Hannun, C.~Case, J.~Casper, B.~Catanzaro, G.~Diamos, E.~Elsen, R.~Prenger,
  S.~Satheesh, S.~Sengupta, A.~Coates \emph{et~al.}, ``Deep speech: Scaling up
  end-to-end speech recognition,'' \emph{arXiv preprint arXiv:1412.5567}, 2014.

\bibitem{panayotov2015librispeech}
V.~Panayotov, G.~Chen, D.~Povey, and S.~Khudanpur, ``Librispeech: an asr corpus
  based on public domain audio books,'' in \emph{Proc. {ICASSP}}.\hskip 1em
  plus 0.5em minus 0.4em\relax IEEE, 2015, pp. 5206--5210.

\bibitem{commonvoice2013}
\BIBentryALTinterwordspacing
Mozilla, ``Common voice,'' 2013. [Online]. Available:
  \url{\url{https://voice.mozilla.org/en/datasets}}
\BIBentrySTDinterwordspacing

\end{thebibliography}

\end{document}